\documentstyle [aps,multicol,psfig]{revtex}
\renewcommand{\narrowtext}{\begin{multicols}{2} \global\columnwidth20.5pc}
\renewcommand{\widetext}{\end{multicols} \global\columnwidth42.5pc}

\begin{document}

\newcommand{\newc}{\newcommand}

\newc{\be}{\begin{equation}}
\newc{\ee}{\end{equation}}
\newc{\ba}{\begin{eqnarray}}
\newc{\ea}{\end{eqnarray}}
\newc{\bea}{\begin{eqnarray*}}
\newc{\eea}{\end{eqnarray*}}
\newc{\D}{\partial}
\newc{\ie}{{\it i.e.} }
\newc{\eg}{{\it e.g.} }
\newc{\etc}{{\it etc.} }
\newc{\etal}{{\it et al.}}
\newcommand{\nn}{\nonumber}

\newc{\ra}{\rightarrow}
\newc{\lra}{\leftrightarrow}
\newc{\no}{Nielsen-Olesen }
\newc{\lsim}{\buildrel{<}\over{\sim}}
\newc{\gsim}{\buildrel{>}\over{\sim}}

\title{Dark Energy and the quietness of the Local Hubble Flow}
\author{M. Axenides$^a$, L. Perivolaropoulos$^b$
}
\address{$^a$ Institute of Nuclear Physics, N.C.R.P.S. Demokritos,  153
10, Athens, Greece \\ $^b$ Department of Physics, University of
Patras, Greece.}

\date{\today}
\maketitle

\begin{abstract}

The linearity and quietness of the Local ($< 10 Mpc$) Hubble Flow
(LHF) in view of the very clumpy local universe is a long standing
puzzle in standard and in open CDM cosmogony. The question
addressed in this paper is whether  the antigravity component of
the recently discovered dark energy can cool the velocity flow
enough to provide a solution to this puzzle. We calculate the
growth of matter fluctuations in a flat universe containing a
fraction $\Omega_X(t_0)$ of dark energy obeying the time
independent equation of state $p_X = w \rho_X$.  We find that dark
energy can indeed cool the LHF. However the dark energy parameter
values required to make the predicted velocity dispersion
consistent with the observed value $v_{rms}\simeq 40km/sec$ have
been ruled out by other observational tests constraining the dark
energy parameters $w$ and $\Omega_X$. Therefore despite the claims
of recent qualitative studies dark energy with time independent
equation of state can not by itself explain the quietness and
linearity of the Local Hubble Flow.

\end{abstract}

\narrowtext

There is mounting evidence\cite{Turner:2001yu} during the past
three years that even though the geometry of the universe is flat,
a large part of its energy density is smooth and exerts a
repelling gravitational force (antigravity) leading to an
accelerating expansion on cosmological scales. This type of energy
density $\rho_X$  (called `dark energy') consists a fraction
$\Omega_X(t_0)\equiv \Omega_{0X}$ of the critical energy density
and has an equation of state\cite{Turner:1998ex} $p_X = w \rho_X$
with $w<-1/3$. The direct evidence for the existence of dark
energy comes from distance measurements\cite{Perlmutter:1998np} of
type Ia supernovae (SNe, Ia) which indicate that the expansion of
the universe is accelerating. Additional evidence comes from the
cosmic microwave background (CMB) anisotropy
measurements\cite{Jaffe:2000tx} which
indicate\cite{Amendola:2000ub,Weller:2001gk} that the fraction of
the total energy density is $\Omega_0 = 1.1\pm 0.07$ while
measurements of the matter density fraction
indicate\cite{Turner:2001yu} that $\Omega_{0M}=0.35\pm 0.07$.
Thus, there is converging evidence that about 2/3 of the total
energy density of the universe is in the form of a smooth
component with negative pressure.

The combination of various observational tests leads to a
constrain for the parameters $\Omega_{0X}$ and $w$
as\cite{Bean:2001xy,Huterer:2000mj,Turner:2001yu}

\be 0.85 \gsim \Omega_{0X}=1-\Omega_{0M} \gsim 0.55 \ee and \be
\label{omran} -0.7 \gsim w \gsim -1 \ee at the 99\% confidence
level. It is therefore important to study additional cosmological
effects of the dark energy in an effort to
\begin{enumerate}
\item
Impose further constraints on $\Omega_{0X}$ and $w$ thus
constraining microphysical theories that predict values for these
parameters
\item
Address cosmological puzzles that may require the features of dark
energy for their solution.
\end{enumerate}

The linearity and quietness of the Local Hubble Flow (LHF) is one
such puzzle\cite{s86}. It is based on three remarkable properties
of the local ($\lsim 10Mpc$) velocity field that are not
consistent \cite{Governato:1996cu} with standard ($\Omega_{0M}
=1$, $\Omega_{0X}=0$) CDM (SCDM)
\begin{enumerate}
\item
The linearity of the velocity-distance relation down to small
distances ($\sim 1.5 Mpc$)
\item
The closeness of the global and local rates of expansion
\item
The small local velocity dispersion around the Hubble law
($\sigma_v \simeq 40 km/sec$ \cite{Ekholm:2001}) \end{enumerate}
It has been proposed\cite{Baryshev:2001xd} based on qualitative
arguments, that the properties of dark energy can be manifest in
the dynamical features of the LHF in such a way as to remedy this
descrepancy with SCDM. The arguments of Ref.
\cite{Baryshev:2001xd} are based on the definition of a critical
distance $r_Q (M,t)$ such that for scales $r>r_Q$ the repulsive
force of dark energy dominates over the gravity of a central mass
concentration $M$ at time $t$. Thus, for $r>r_Q$ the total
gravitating mass \be M_{tot} = {{4\pi}\over 3} (1 +3w) \rho_X r^3
+ M \ee becomes negative. According to Ref.
\cite{Baryshev:2001xd}, at any time $t$, the dynamics of all
scales $r>r_Q$ are dominated by the effects of dark energy.
Therefore if a galaxy in the neighborhood of the Local Group, had
spent enough time beyond $r_Q$  could have its peculiar velocity
adiabatically cooled enough to be consistent with the observed
cold LHF. The arguments of Ref. \cite{Baryshev:2001xd} are
qualitative because they are based on an abrupt transition from
matter dominated to dark energy dominated scales and no attempt is
made to calculate the total velocity growth factor in a two
component cosmological setup. Nevertheless, the conclusion of Ref.
\cite{Baryshev:2001xd} is that the adiabatic cooling during dark
energy domination in the neighborhood of the Local Group is enough
to cool the matter induced peculiar velocities to levels
consistent with observations\cite{Ekholm:2001} for $w=-2/3$,
$\Omega_X \simeq 0.7$.

The main goal of this paper is to quantify this proposal by
calculating the growth of density and velocity fluctuations in a
flat universe ($\Omega_{0X}+ \Omega_{0M} =1$) containing dark
energy characterized by $\Omega_{0X}$ and constant $w$. This
calculation leads to suppression growth factors for velocity and
density perturbations that depend on $\Omega_{0X}$ and $w$. The
main question we wish to address using this calculation is the
following: Are there values of $\Omega_{0X}$ and $w$ consistent
with present constraints that can produce sufficient growth
suppression leading to the observed linear and quite LHF?

We thus consider a flat, two component cosmological model with
matter ($p_M = 0$) and dark energy ($p_X = w \rho_X$). The
Friedman equation describing the evolution of the scale factor is
\be \label{freq} ({\dot{a}\over a})^2=H_0^2 [\Omega_{0M} ({a\over
a_0})^{-3} + (1-\Omega_{0M}) ({a\over a_0})^{-3(1+w)}] \ee where
we have used the fact that $\rho_X \sim a^{-3(1+w)}$ (a
consequence of the continuity equation $T^\nu_{\mu;\nu}=0$).
Setting $a_0 \equiv a(t_0) = 1$ equation (\ref{freq}) takes the
form of
 \be \label{freq1} \dot{a}^2 = H_0^2 [\Omega_{0M} a^{-1} +
(1-\Omega_{0M})a^{-1-3w}] \ee

The equations describing the evolution of density and velocity
fluctuations in this cosmological model can be obtained as usual
from the continuity (for density), momentum (for velocity) and
Poisson (for gravitational potential) equations \cite{lss}.
Combining these equations and linearizing with respect to the
gravitational potential, density and velocity perturbations we
obtain a second order linear differential equation for the density
contrast \be \label{dc} \delta \equiv {{\rho_M (x,t) -
\bar{\rho_M}(t)}\over {\bar{\rho_M}(t)}} \ee as \be \label{fleq}
{{\partial^2 \delta}\over {\partial t^2}} + 2 {\dot{a} \over a}
{{\partial \delta}\over {\partial t}}=4\pi G \bar{\rho_M}(t)
\delta = {{3 H_0^2 \Omega_{0M}}\over {2 a^3}}\delta \ee For the
corresponding peculiar velocity \be \label{vpec} \vec{v}_{pec}
\equiv a {{d \vec{x}}\over {dt}} = {{d \vec{r}}\over {dt}}-
{\dot{a}\over a} \vec{r} \ee (where $\vec{x} = {\vec{r} \over a}$
is the comoving coordinate) we obtain \be \label{vleq}
\vec{v}_{pec}(\vec{x},t)= \dot{a} a {{\partial \delta}\over
{\partial a}} f(\vec{x}) \ee where $f(\vec{x})$ depends on the
initial spatial dependence of density fluctuations. Using equation
(\ref{freq1}) to change variables from $t$ to $a$, equation
(\ref{fleq}) becomes \ba \label{fleq1} &&a^2 [\Omega_{0M} +
(1-\Omega_{0M})a^{-3w}]{{\partial^2}\over {\partial a^2}}\delta +
\nn \\  && a [{{3 \Omega_{0M}}\over 2} +
3(1-\Omega_{0M})(1-{{1+w}\over 2}) a^{-3w}]{{\partial \delta}\over
{\partial a}}  \nn \\ && ={{3 \Omega_{0M}}\over 2} \delta \ea The
general solution of this equation is a superposition of two modes
a growing mode $\delta_+$ and a decaying mode $\delta_-$. Our
numerical solution to this equation will be dominated by the
growing mode.

Equation (\ref{fleq1}) may now be solved numerically with
appropriate initial conditions to obtain the growth factors for
density and velocity fluctuations evolved from \be \label{ai}
a_i=a_{eq}={{4.31 \times 10^{-5}}\over {h^2 \Omega_{0M}}} \ee to
the present $a_0 = 1$. Assuming an initially small density
fluctuation we may set \be \label{ic1} \delta(t_i \simeq 0) = 0
\ee while since matter dominates over dark energy at early times
we may set \be \label{ic2} {{\partial \delta}\over {\partial a}}
(t_i \simeq 0) = 1 \ee since $\delta_+ (a) \sim a$ in a flat
matter dominated universe.

In the above initial conditions we have ignored spatial dependence
and possible constant factors multiplying $\delta$ since we are
not interested in the absolute values of fluctuations but in the
growth factors \ba g_d(\Omega_{0M} ,w) & \equiv & {{\delta (t_0,
\Omega_{0M}, w)}\over {\delta (t_{eq}, \Omega_{0M}, w)}} \\
g_v(\Omega_{0M} ,w) & \equiv & {{v_{pec} (t_0, \Omega_{0M},
w)}\over {v_{pec} (t_{eq}, \Omega_{0M}, w)}} \ea We have solved
equation (\ref{fleq1}) with the initial conditions (\ref{ic1}) and
(\ref{ic2}) and we have constructed contour plots for the relative
growth factors ${{g_d(\Omega_{0M} ,w)}\over {g_d(1 ,0)}}$ and
${{g_v(\Omega_{0M} ,w)}\over {g_v(1 ,0)}}$ (Figs 1 and 2). Even
though each individual growth factor depends on the value of $h$
the {\it relative} growth factors as defined above are independent
of the Hubble constant.
\begin{figure}
   \psfig{file=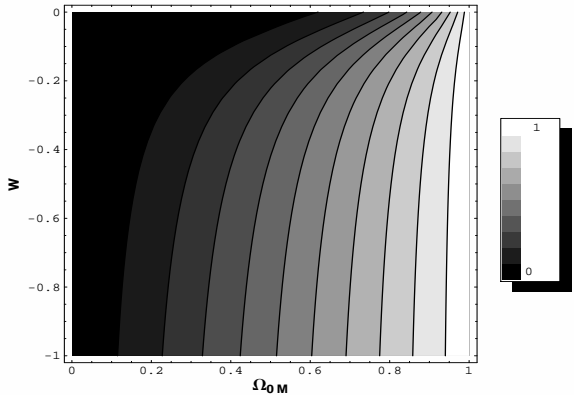,width=10.0cm,height=5.5cm,angle=0}
   \vspace{15pt}
   \caption{The relative growth factor ${{g_d(\Omega_{0M} ,w)}\over {g_d(1,0)}}$ for density
   fluctuations for a flat cosmological model as a function of
   the cosmological parameters $\Omega_{0M}$ and $w$. For small
   $\Omega_{0M}$ and large $w$ the growth gets suppressed as
   expected due to shorter domination of matter.}
   \label{fig:denpert}
\end{figure}
These plots indicate that there is a suppression of growth for
fluctuations which as discussed in the introduction increases with
$w$ and $\Omega_{0X}$. Similar linear growth suppression factors
have also been discussed elsewhere \cite{Fry:zy} using both
numerical and analytical methods and they affect the overall
normalization of the power spectrum today. This normalization is
expressed by $\sigma_8$, the rms amplitude of matter perturbations
on a scale of $8h^{-1} Mpc$. Given a COBE normalized scale
invariant cosmological model $\sigma_8$ decreases with $w$,
dropping significantly\cite{Huterer:2000mj} below its presently
observed value $\sigma_8 = (0.56\pm0.1) \Omega_{0m}^{-0.47}$ for
$w>-0.7$ ($\Omega_{0m}\simeq 0.3$). This constraint which becomes
slightly stronger for smaller $\Omega_{0m}$ has effectively ruled
out values of $w>-0.7$ as shown in equation (\ref{omran}). Recent
studies\cite{recent-s8} indicating a lower value of $\sigma_8$
could relax somewhat the constraint (\ref{omran}) towards larger
values of $w$.

The deviation from a pure Hubble flow characterized by the
observed radial peculiar velocity dispersion is
measured\cite{Ekholm:2001} to be \be \label{vmeas} v_{rms}\sim 40
km/sec \ee  An important challenge for cosmological models is the
establishment of their consistency with such low velocity
dispersion for parameter values that are compatible with other
observations. High resolution CDM N body
simulations\cite{Governato:1996cu} have shown that such low
velocity dispersion is incompatible with both standard CDM (SCDM,
$\Omega_{0m}=1$, $h=0.5$, $\sigma_8 = 0.7$) and open CDM
(OCDM,$\Omega_{0m}=0.3$, $\Omega_{0X}=0$, $h=0.75$, $\sigma_8 =
1$) which according to the above simulations typically predict
velocity dispersions \ba \label{vsim} v_{rms} (SCDM)& \sim &
300-700km/sec \nn
\\ v_{rms} (OCDM)& \sim & 150-300km/sec \ea respectively.
These simulations have a Harrison-Zeldovich spectrum ($n=1$) and
their conclusions are stated not to sensitively depend on $n$.
According to these simulations, neither of these models can
produce a single candidate LG with the observed velocity
dispersion in a volume $10^6 Mpc^3$.
\begin{figure}
   \psfig{file=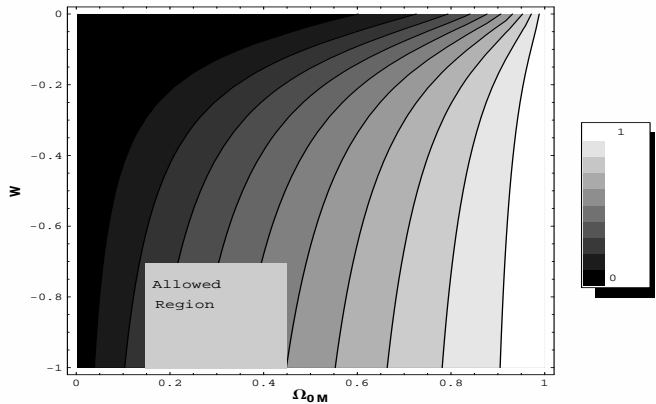,width=10.0cm,height=5.5cm,angle=0}
   \vspace{15pt}
   \caption{The relative growth factor ${{g_v(\Omega_{0M} ,w)}\over {g_v(1,0)}}$   for velocity
   fluctuations for a flat cosmological model as a function of
   the cosmological parameters $\Omega_{0M}$ and $w$.}
   \label{fig:velpert}
\end{figure}
Thus using eq. (\ref{vsim}) and eq. (\ref{vmeas}) we conclude that
a suppression factor \be \label{sfr} {\bar g}_v\equiv
{{g_v(\Omega_{0M},w)}\over {g_v(1,0)}} \in [0.06,0.13] \ee  over
the SCDM predicted velocity dispersion is required to make the
model predictions consistent with observations. Inspection of the
suppression factor contour plot of Fig. 2 indicates that the
presence of dark energy can not by itself provide a solution to
this problem. The required values of the suppression factor are
achieved within the first and part of the second darkest stripes
of Fig. 2. However, as shown in Fig. 3, no part of this parameter
space is consistent with constrains obtained from other
observational tests which seem to indicate\cite{Bean:2001xy} that
$\Omega_{0M}\in [0.15,0.45]$ and $w\in [-1,-0.7]$ at 99\% level.
This result is independent of $h^2$, $\sigma_8$ and $n$ since the
evaluation of ${\bar g}_v$ is insensitive to these parameters.

We conclude that the presence of dark energy with time independent
equation of state can not by itself resolve the puzzle of the LHF
linearity and quietness by suppressing the growth of velocity
fluctuations. The required velocity dispersion suppression  is
achieved for parameter values that are incompatible with other
observational tests and requires values of $\Omega_{0M}$ that are
too low (e.g. $w\lsim -0.7$ leads to $\Omega_{0M} \simeq 0.1$) out
of the currently allowed interval. Alternatively for a fixed value
of $\Omega_{0M} \simeq 0.3$, the resolution of the LHF puzzle with
dark energy requires too large $w$ ($w\gsim -0.3$).

\begin{figure}
   \psfig{file=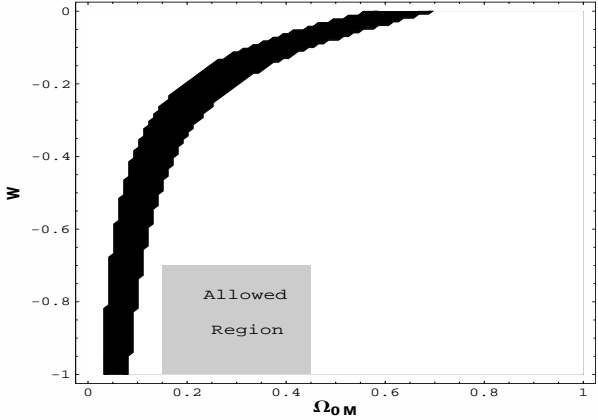,width=10.0cm,height=5.5cm,angle=0}
   \vspace{15pt}
   \caption{The allowed parameter region based on observational constraints (grey square)
   compared to the parameter region (eq. (\ref{sfr})) required to resolve the LHF quietness puzzle (dark stripe).}
   \label{fig:constr}
\end{figure}

An interesting extension of our work could be the introduction of
a time dependence in the dark energy equation of state along the
lines of quintessence models \ie $ p=(w_0 + w_1 z)\rho$. This type
of equation of state allows for non-trivial effects of dark energy
even at early times which could lead to a similar suppression of
velocity fluctuation growth at higher values of $\Omega_{0M}$
and/or lower values of $w$ thus being consistent with current
observational constrains.

{\bf Acknowledgements:} LP acknowledges support by the University
of Patras through the `Karatheodoris' research grant No. 2793.
This work is the result of a network supported by the European
Science Foundation.

\widetext
\end{document}